\documentclass[prd,aps,superscriptaddress,nofootinbib,showpacs,twocolumn]{revtex4}

\usepackage[intlimits]{amsmath}
\usepackage{amsfonts}
\usepackage{amssymb,amscd}
\usepackage{color}
\usepackage[dvips]{epsfig}

\newcommand{\beq}{\begin{equation}}
\newcommand{\eeq}{\end{equation}}
\newcommand{\beqa}{\begin{eqnarray}}
\newcommand{\eeqa}{\end{eqnarray}}
\newcommand{\be}{\begin{equation}}
\newcommand{\ee}{\end{equation}}

\newcommand{\vp}{\vec{p}}
\newcommand{\vq}{\vec{q}}
\newcommand{\vk}{\vec{k}}
\newcommand{\vrr}{\vec{r}}
\newcommand{\op}{\omega_{p}}
\newcommand{\oq}{\omega_{q}}
\newcommand{\ok}{\omega_{k}}
\newcommand{\orr}{\omega_{r}}
\newcommand{\intq}{\sum_{n_q} \int \frac{d^3q}{(2\pi)^3}}
\newcommand{\gS}[1]{#1\!\!\!\!\!\not~}

\newcommand{\pslash}{\gS{p}}

\begin{document}

\title{On critical scaling at the QCD $N_f=2$ chiral phase transition}

\author{Christian~S.~Fischer}
\affiliation{Institut f\"ur Theoretische Physik,
  Justus-Liebig-Universit\"at Gie\ss{}en,
  Heinrich-Buff-Ring 16,
  D-35392 Gie\ss{}en, Germany}
\affiliation{GSI Helmholtzzentrum f\"ur Schwerionenforschung GmbH,
  Planckstr. 1  D-64291 Darmstadt, Germany.}
\author{Jens~A.~Mueller}
\affiliation{Institut f\"ur Kernphysik,
  Technische Universit\"at Darmstadt,
  Schlossgartenstra{\ss}e 2,\\
  D-64289 Darmstadt, Germany}

\date{\today}
\begin{abstract}
We investigate the critical scaling of the quark propagator of 
$N_f=2$ QCD close to the chiral phase transition at finite temperature. 
We argue that it is mandatory to take into account the back-reaction 
effects of pions and the sigma onto the quark to observe critical 
behavior beyond mean field. On condition of self-consistency of the
quark Dyson-Schwinger equation we extract the scaling behavior 
for the quark propagator analytically. Crucial in this respect is 
the correct pion dispersion relation when the critical temperature is 
approached from below. Our results are consistent with the 
known relations for the quark condensate and the pion decay constant
from universality. We verify the analytical findings also numerically 
assuming the critical dispersion relation for the Goldstone bosons.

\end{abstract}

\maketitle

\section{Introduction}

The phase diagram of strongly interacting quark matter 
has received a lot of attention, both from theoretical 
and experimental perspectives. 
One of the debated issues in this respect is the nature 
of the phase transition in $N_f=2$ QCD in the chiral 
limit. It has long been suggested that the 
$SU(2)_V \times SU(2)_A \simeq O(4)$ symmetry in this case
is broken spontaneously to $SU(2)_V \simeq O(3)$ by 
a second order chiral phase transition of the
$O(4)$-universality class in 3-dimensions
\cite{Pisarski:1983ms}. The underlying symmetry breaking pattern 
results from the assumption that the anomalously broken
$U_A(1)$ symmetry of QCD leads to a relatively heavy pseudoscalar
flavor singlet meson also for temperatures around $T_c$.

One of the tools to investigate this issue is lattice 
gauge theory, see e.g. \cite{Karsch:2007dp,Philipsen:2008gf}. 
The problem is delicate 
since the analysis of a potential second order phase 
transition in a finite volume is elaborate and light 
quarks are expensive in lattice computations. As a result
there are indications for $O(4)$ scaling from 
simulations with Wilson fermions \cite{Ali Khan:2000iz}
and very recently from results with staggered fermion actions 
\cite{Ejiri:2009ac,Karsch:2010ya,Kaczmarek:2011zz}. However,
since there are also arguments in favor of a first order 
transition \cite{D'Elia:2005bv,Chandrasekharan:2007up}, the
problem seems not yet completely settled.

The symmetry breaking pattern with an anomalously broken
$U_A(1)$ symmetry is a widely used assumption for the construction 
of effective models as for example the Polyakov-loop quark meson 
model (PQM), see e.g. 
\cite{Schaefer:2007pw,Braun:2010vd,Skokov:2010uh,Herbst:2010rf}. 
Recent constructions of effective models 
include and investigate the effects of the anomaly 
\cite{Schaefer:2008hk,Chen:2009gv}. The issue of critical
scaling has been addressed in \cite{Berges:1997eu,Bohr:2000gp}.

A third approach to the phase diagram are functional methods 
like the functional renormalization group 
\cite{Braun:2006jd,Braun:2007bx,Braun:2009gm} or
Dyson-Schwinger equations (DSEs) \cite{Fischer:2009wc,Fischer:2011mz,Fischer:2010fx} 
applied directly to QCD. In principle, these methods are capable to  
link information on the microscopic degrees of freedom of the theory 
to the characteristics of the phase transitions. Recent 
works include the calculation of the nonperturbative Polyakov-loop 
potential from the propagators of Yang-Mills theory
\cite{Braun:2007bx}, the investigation of the chiral and deconfinement 
transition \cite{Braun:2009gm,Fischer:2011mz,Fischer:2010fx},
quark spectral functions \cite{Mueller:2010ah} and the exploration 
of color superconducting phases \cite{Nickel:2006vf}.

Although DSEs are well established they are usually not used to study 
critical phenomena and it may be even unclear to what extend they are 
suitable. In fact, up to now critical scaling in DSEs has only been 
observed on the mean field level \cite{Roberts:2000aa,Blank:2010bz}. 
The main goal of this paper is to go beyond these studies. We employ 
the Dyson-Schwinger equations to explore the $O(4)$ scaling in the 
quark propagator at the chiral phase transition for $N_f=2$ QCD.
By including the effects of pion and sigma back-reaction onto the 
quarks along the lines of the zero temperature studies in 
Refs.~\cite{Fischer:2008wy,Fischer:2008sp} we identify a mechanism 
that generates critical scaling beyond the mean field level. Our 
results therefore provide a basis for further studies of the chiral 
phase transition in this framework. They also serve to study the 
connection between the scaling properties in the effective theory 
and the behavior of the fundamental microscopic degrees of freedom 
such as the quark propagator. 

The paper is structured as follows: In section \ref{sec:qDSE_Goldstone}
we discuss our approximation scheme for the DSE of the quark propagator 
at finite temperature. We then analyze the scaling behavior of the DSEs
analytically in section \ref{sec:scaling} and demonstrate that the 
mechanism for the onset of critical scaling beyond the mean field 
level in the quark propagator is located in the pion and sigma dispersion 
relations. For our numerical investigations we discuss a tractable 
truncation scheme in section \ref{sec:trunc} and verify our analytical 
findings in section \ref{sec:o4numerics} under the assumption of a 
scaling law for the pion velocity. We conclude and summarize in section
\ref{sec:con}.

\section{Quark Dyson-Schwinger equation
 and effects from Goldstone bosons\label{sec:qDSE_Goldstone}}

\subsection{Backcoupling mesons onto quarks}

\begin{figure}[ht]
\centerline{\epsfig{file=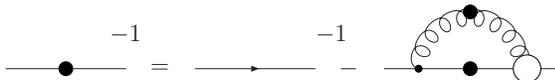,width=0.9\columnwidth}}
\caption{The Schwinger-Dyson equation for the fully dressed
quark propagator}\label{fig:dse1}
\end{figure}

The DSE for the quark propagator is given diagrammatically in 
Fig.~\ref{fig:dse1}. The unknown quantities in this 
equation are the dressed gluon propagator and the one-particle 
irreducible (1PI) quark-gluon vertex. We are particularly interested in 
the behavior of the quarks close to the chiral limit and in the 
vicinity of the chiral phase transition. There the relevant degrees of 
freedom are associated with long-range correlations and can be identified 
as the Goldstone modes and the associated radial excitation. 
These contribute to the DSE of the quark-gluon vertex as outlined in 
Ref.~\cite{Fischer:2008wy}. We first briefly summarize the relevant 
findings of Ref.~\cite{Fischer:2008wy} and subsequently
consider their implications at finite temperature.

\begin{figure}[t!]
\vspace*{5mm}
\centerline{\epsfig{file=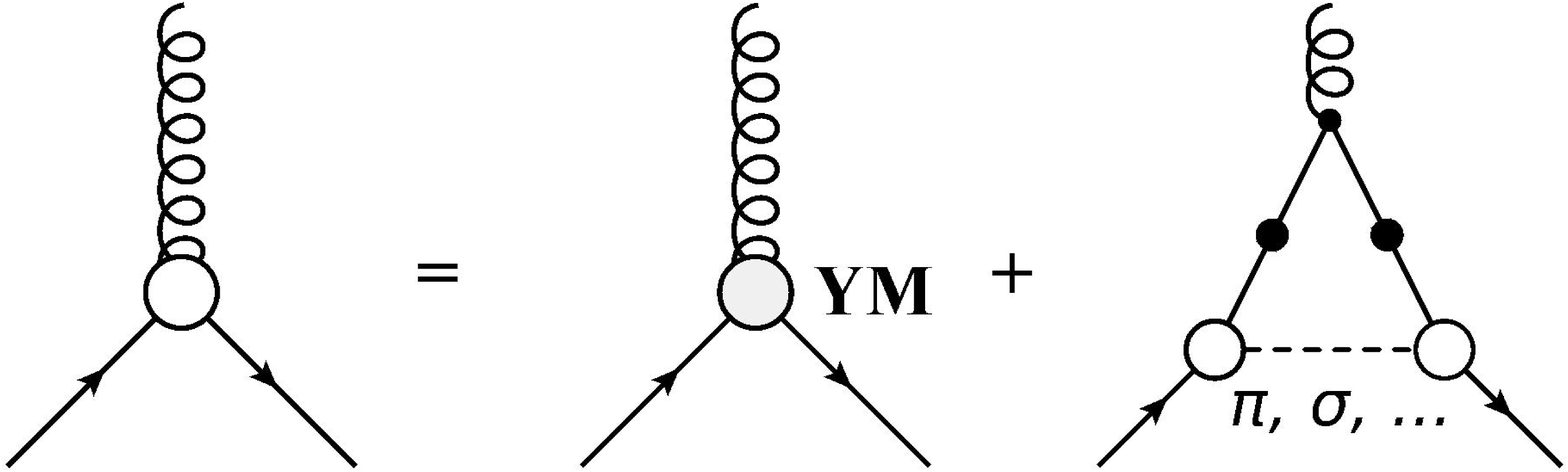,width=.65\columnwidth}}
\begin{Large}$\downarrow$\end{Large}
\centerline{\epsfig{file=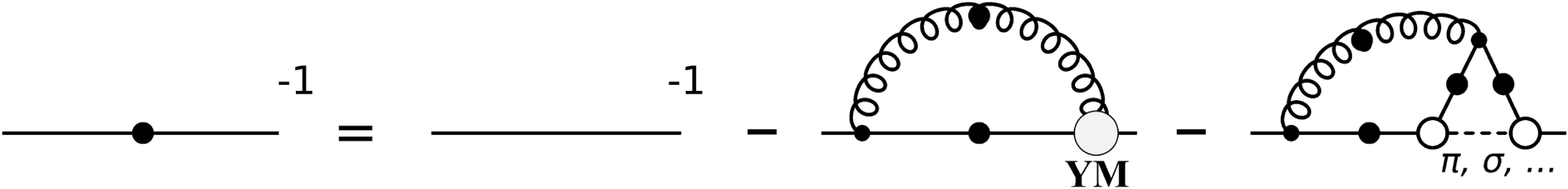,width=1.\columnwidth}}
\caption{The Dyson-Schwinger equation for the quark propagator
using a quark-gluon vertex that includes back-coupling effects from 
pions and the sigma meson. \label{fig:quarkdse}}
\end{figure}

The DSE of the 1PI quark-gluon vertex can be analyzed
in terms of contributions from pure Yang-Mills theory and unquenching
effects in the form of the back-coupling of meson degrees of freedom
onto the quark-gluon interaction \cite{Fischer:2008wy,Fischer:2008sp}.
To lowest order in a skeleton expansion these appear as
one-meson exchange between quark and anti-quark
as displayed in the upper panel of Fig.~\ref{fig:quarkdse}. 
Since we are primarily interested in long-range meson contributions, 
we need to keep this term explicitly in our approximation scheme of 
the quark-gluon vertex.

At zero temperature and at physical quark masses the meson back-reaction 
term is dominated by the exchange of the lightest mesons, i.e. pions. 
At finite temperature and close to the chiral limit the $SU(2)_A$-symmetry 
gets restored at the critical temperature $T_c$ and the order parameter 
correlator needs to be taken into account. This is done by considering the 
flavor singlet sigma meson as well. The pseudoscalar flavor singlet meson,
however, is assumed to receive a topological mass due to anomalous 
non-conservation of the $U_A(1)$-symmetry and is therefore suppressed.
Consequently, we approximate the quark-gluon vertex by a sum of purely 
gluonic contributions not affected by critical modes and the meson exchange 
contribution containing pions and the sigma meson. The resulting DSE
for the quark propagator is shown in the lower panel of Fig.~\ref{fig:quarkdse}. 
 
The calculation of the quark self-energy with meson exchange is in general 
quite demanding. It contains a two-loop diagram containing the full Bethe-Salpeter 
amplitudes, $\Gamma_{\pi,\sigma}$, which can in principle be determined 
from the corresponding Bethe-Salpeter equation. In order to obtain a tractable 
truncation that preserves the long range effects close the chiral phase 
transition we will use some low-energy properties of pions in the next section.

\subsection{The DSEs at finite temperature \label{sec:t}}
Before we go into the details, we first give the formal expression for the DSE of the
quark propagator as shown in Fig.~\ref{fig:quarkdse}. At 
finite temperature it is given by 
\begin{widetext}
\begin{eqnarray}
S^{-1}(\vp,\op)&=&Z_2\,S^{-1}_{0}(\vp,\op)+\Sigma_{\text{YM}}(\vp,\op)+\Sigma_{\pi}(P)+\Sigma_{\sigma}(\vp,\op)
\label{eq:DSE}
\end{eqnarray}
where the self-energies are
\begin{eqnarray}
\Sigma_{\text{YM}}(\vp,\op)&=&Z_{1F}\,g^2 \,C_F \,T\,\sum_{n_q}\int\frac{d^3q}{(2\pi)^3}\; \gamma_{\mu}\,S(\vq,\oq)
\,\Gamma^{\text{YM}}_{\nu}(\vq,\oq,\vp,\op)\,D^{}_{\mu\nu}(\vk,\ok)\,\label{eq:YM}\\[.3cm]
\Sigma_{\pi}(\vp,\op)&=&2Z_{1F}\,g^2 \,C_F \,T\,\sum_{n_q}\int\frac{d^3q}{(2\pi)^3}\; \gamma_{\mu}\,S(\vq,\oq)
\,\Gamma^{\pi}_{\nu}(\vq,\oq,\vp,\op)\,D^{}_{\mu\nu}(\vk,\ok)\;\; \label{eq:pi_meson}\\[.3cm]
\Sigma_{\sigma}(\vp,\op)&=&Z_{1F}\,g^2 \,C_F \,T\,\sum_{n_q}\int\frac{d^3q}{(2\pi)^3}\; \gamma_{\mu}\,S(\vq,\oq)
\,\Gamma^{\sigma}_{\nu}(\vq,\oq,\vp,\op)\,D^{}_{\mu\nu}(\vk,\ok)\, \label{eq:sig_meson}
\end{eqnarray}
\end{widetext}
with $k=p-q=(\vk,\ok)$, the temperature $T$, the Casimir
$C_{F}=(N_{c}^{2}-1)/(2N_{c})$ and the renormalization 
factors $Z_{1F}$ of the quark gluon vertex and $Z_2$ of 
the quark propagator.
We used the abbreviations $\Gamma_\nu^{(\pi,\sigma)}$ for 
the triangle meson exchange that is part of the quark-gluon 
vertex. This should not be confused with the Bethe-Salpeter 
amplitudes $\Gamma_{\pi,\sigma}$.
All color and flavor traces have 
been carried out already\footnote{Note that the flavor trace 
for the self-energy part due to the pion backreaction results in
a factor of two, since the flavor of the external quark is fixed. 
This has been overlooked in Refs.~\cite{Fischer:2008wy,Fischer:2008sp}, 
where a factor of three has been used leading to a slight 
overestimation of the pion backcoupling effects.}. 
In this work we only consider 
$N_f=2$ and $N_c=3$. The gluon propagator 
is denoted by $D_{\mu\nu}$ and the purely gluonic part of 
the dressed quark-gluon vertex by $\Gamma_{\nu}^{\text{YM}}$.
Using a decomposition in Dirac components the
dressed quark propagator can be parametrized by
\begin{eqnarray}
S(\vp,\op) \!\!&=& \!\!\left[-i \gamma_4 \op \,C(\vp,\op) 
- i \vec{\gamma}\vp \,A(\vp,\op) + B(\vp,\op)\right]^{-1} \nonumber\\
&=& i \gamma_4 \op \,\sigma_C(\vp,\op) +  
i \vec{\gamma}\vp \,\sigma_A(\vp,\op) + \sigma_B(\vp,\op) \nonumber\\
\end{eqnarray}
with the scalar dressing function $B(\vp,\op)$ and the two 
vector dressing functions $A(\vp,\op)$ and $C(\vp,\op)$. 
Its bare counterpart reads 
$S^{-1}_{0}(\vp,\op) = -i \pslash + m$ and we work with
very small quark masses $m \rightarrow 0$.

The details of the non-perturbative
gluon propagator and Yang-Mills quark-gluon vertex are not 
important for the scaling behavior in the vicinity of 
the chiral phase transition. 
This may be expected from the universality argument and will
also show up in our analysis presented in the next section.
In this section we only state the general tensor decomposition
of the gluon propagator and Yang-Mills quark-gluon vertex
that is used in this paper. For further details we refer 
to section \ref{sec:trunc} where we consider the numerical solution.
Throughout this paper we will work in Landau gauge. 
In this case the most general representation of the
gluon propagator at finite temperature is composed of
 two dressing functions. It can be written as
\begin{eqnarray}
D_{\mu\nu}(\vk,\ok) &=& \frac{Z_T(\vk,\ok)}{k^2} P_{\mu\nu}^T 
+\frac{Z_L(\vk,\ok)}{k^2} P_{\mu,\nu}^L
\end{eqnarray} 
with transverse and longitudinal projectors ($i,j=1 \dots 3$)
\begin{eqnarray}
P_{\mu\nu}^T(\vk,\ok) &=& 
   \left(\delta_{i,j}-\frac{k_i k_j}{\vk^2}\right) 
   \delta_{i\mu}\delta_{j\nu}\,, \nonumber\\
P_{\mu\nu}^L(\vk,\ok) &=& P_{\mu\nu}(\vk,\ok) - P_{\mu\nu}^T(\vk,\ok) \,.
\end{eqnarray} 
and dressing functions $Z_T(\vk,\ok)$ and $Z_L(\vk,\ok)$.
For the Yang-Mills part of the quark-gluon vertex we adopt
the rainbow ladder approximation
\begin{equation}
 \Gamma_{\nu}^{\text{YM}}(\vq,\oq,\vp,\op) \rightarrow \gamma_\nu \Gamma^{\text{YM}}(\vk,\ok)
\end{equation}
where $\Gamma^{\text{YM}}(\vk,\ok)$ is a vertex model depending
only on the gluon momentum. 

Now we need to specify the meson contributions $\Gamma^{\pi,\sigma}_{\nu}$ in
Eqs.(\ref{eq:pi_meson}) and (\ref{eq:sig_meson}). These are given by
\begin{eqnarray}
  \Gamma^{\pi,\sigma}_{\nu}&=&\!\!T\sum_{n_r} \int\frac{d^3r}{(2\pi)^{3}} \,
  \Gamma^{}_{\pi,\sigma}(\vq,\oq,\vrr,\orr)S(\vq+\vrr,\oq+\orr)
\nonumber\\
&& \hspace*{-2mm}\times  \gamma_\nu S(\vp+\vrr,\op+\orr)\Gamma^{}_{\pi,\sigma}(\vp,\op,\vrr,\orr)D_{\pi,\sigma}(\vrr,\orr) \nonumber\\
\label{eq:meson}
\end{eqnarray}
with the Bethe-Salpeter amplitudes $\Gamma_{\pi,\sigma}$ and the
meson propagators $D_{\pi,\sigma}$. In the following we consider
the pion Bethe-Salpeter amplitude and propagator.

In Ref.~\cite{Son:2001ff} Son and Stephanov showed that the
real part of the pion dispersion relation in the symmetry broken
phase is given by
\begin{equation}
 \omega^2=u^2(\vp^2+m^2_\pi)\,
\end{equation}
where $u$ denotes the pion velocity and $m_\pi$ the 
pion screening mass. At $\vp=0$ the energy of a pion
$E_\pi=u\,m_\pi$ is called the pion pole mass.
Furthermore as already noted in Ref.~\cite{Pisarski:1996mt}
there are two distinct pion decay constants in the medium.
At finite temperature these are given by $f_s$ transverse
to the heat bath and $f_t$ longitudinal to the heat bath.
The ratio of both determines the 
pion velocity
\beq
u^2 = \frac{f^2_s}{f^2_t}\,.
\eeq 
At temperature $T=0$ we have $f_\pi = f_s = f_t$ and
consequently $u=1$, whereas at nonzero temperature this 
is no longer the case.
We stress that the pion decay constants constitute
static quantities and are in principle calculable 
in a thermodynamic equilibrium approach as pointed out in \cite{Son:2001ff}. 
The relation between the pion decay constant,
the full quark propagator and the pion Bethe-Salpeter amplitude
in the vacuum is outlined in Ref.~\cite{Maris:1997hd}.
To obtain the decay constants at finite temperatures
we may proceed along the same lines considering
the in-medium pion propagation 
\begin{equation}
 D_\pi = \frac{1}{\omega_p^2 + u^2 (\vp^2 + m_\pi^2)}\,.
\end{equation}
This leads to the following expression for the decay constants
\begin{equation}
 \tilde{P}_{\mu}f_t
=\,3\,\textrm{\large{tr}}_{D}T\sum_{n_q}\int\frac{d^3q}{(2\,\pi)^3}\,\Gamma_\pi(q,P)\,
S(q+P)\gamma_5\gamma_{\mu}S(q) \label{pi_decay}
\end{equation}
valid on the pion mass shell. We used the 
abbreviation $\tilde{P}_{\mu}=(u\vec{p},\op)$.
In the chiral limit $m_\pi^2=0$ and the expression 
evaluated in the limit $P_\mu\rightarrow 0$
determines the decay constants: The longitudinal
decay component $f_t$ is obtained from the time component
of Eq.~(\ref{pi_decay}), whereas the transverse component 
$f_s=u f_t$ can be extracted from the spatial components.
In order to be able to compute them  
and to specify the pion contribution
$\Gamma^{\pi}_\nu$ we need the pion Bethe-Salpeter 
amplitude $\Gamma_\pi$. It satisfies a homogenous 
Bethe-Salpeter equation which in principle needs to be solved at finite 
temperature. This is a tremendous task beyond the scope of 
the present work. However from the axial vector Ward-Takahashi identity
in the chiral limit it follows that the Bethe-Salpeter amplitude for $P^2=0$ 
may be written as
\begin{equation}
 \Gamma_\pi(q,0)=\gamma_5\frac{B(q)}{f_t}\,. \label{eq:BSA1}
\end{equation}
This constitutes a Goldberger-Treiman like
relation for the quark-pion coupling and
is known to represent
 the leading term also for $P^2 \neq 0$ at zero temperature. 
In the following we approximate the Bethe-Salpeter
amplitude for general momenta by
\begin{equation}
\Gamma_\pi(q,P)\approx \gamma_5 \frac{B(q)}{f_t}\,. \label{eq:BSA2}
\end{equation}
For our purposes this approximation
seems particularly reasonable since it incorporates
the long-range interaction in Eq.~(\ref{eq:meson}) which is important
for the critical behavior.
Using (\ref{eq:BSA2}) in (\ref{pi_decay})
yields the Pagel-Stokar approximation at finite temperature
for the pion decay constants. 

Following the above considerations
and knowledge of the gluon and Yang-Mills quark-gluon vertex dressing functions
provided we obtain a closed system of 
equations for the quark propagator. This truncation also includes 
effects of pion backcoupling.
Up to now we did not consider 
the contribution of the scalar flavor singlet meson $\Gamma^{\sigma}_\nu$. 
We know that its correlator is degenerate
with the $\pi$-correlator for momenta large compared to its mass.
Near the transition temperature and close to the chiral limit
$m \rightarrow 0$ the mass $m_\sigma$ tends to zero
and therefore sufficiently close to $T_c$
we expect the above relations to hold also
for the sigma meson, with $\gamma_5$ in (\ref{eq:BSA2}) replaced by one. 
This will be exploited in the following.

\section{Analytical scaling analysis \label{sec:scaling}}
Having outlined general properties of pions in the symmetry broken phase we 
will now perform a scaling analysis of the quark DSE given in Eq.~(\ref{eq:DSE})
for $T<T_c$ close to the critical temperature.
 We concentrate on the scalar dressing function $B(\vp,\op)$ and work
 with very small but nonzero bare quark mass $m$. 
The reduced temperature is given by $t=(T_c-T)/T_c$. 

The scaling behavior of the pion velocity and decay constants
close to the critical temperature
can be obtained from a matching of an effective theory with
QCD at the scale $m_\sigma$. 
As shown by Son and Stephanov 
in Ref.~\cite{Son:2001ff} this matching reveals
\be
u\sim f_s\sim t^{\nu/2}\;
 \label{scal1}
\ee
using the known scaling relations for the inverse correlation
length and the order parameter 
\be
m_\sigma\sim t^{\nu}\;,\quad\quad \langle \bar{\psi}\psi \rangle\sim t^{\nu/2(1-\eta)}\,.\label{eq:mscaling}
\ee
In $d=3$ dimensions the values for the exponents of the $O(4)$-universality class are
$\nu\approx 0.73$ and $\eta\approx 0.03$, see e.g. \cite{Baker:1977hp,Rajagopal:1992qz}.

For our scaling analysis we assume the scalar function to fulfill a scaling law,
$B\sim t^{x}$,  with some unknown exponent $x$.
On the other hand there is no indication for the vector dressing functions $A$ and $C$ 
to scale with reduced temperature.
The main idea is to utilize self-consistency of the Dyson-Schwinger
equation. This requires left- and right-hand side of (\ref{eq:DSE})
to scale with $t$ in the same way.
For simplicity we ignore the anomalous dimension $\eta$
which would be only a small quantitative correction anyway.
Focusing on $B$ it remains to analyze the scalar projection of the self-energies
given in Eqs.~(\ref{eq:YM}), (\ref{eq:pi_meson}) and (\ref{eq:sig_meson}):
\be
B(t)=\Sigma^{B}_{\text{YM}}(t)+\Sigma^B_{\pi}(t)+\Sigma^B_{\sigma}(t)\,, \label{scalarDSE}
\ee
with $\Sigma^{B}_{\text{YM}}=\text{Tr}(\Sigma_{\text{YM}})/4$ respectively 
$\Sigma^{B}_{\pi,\sigma}=\text{Tr}(\Sigma_{\pi,\sigma})/4$.
The important quantities of the analysis are the scalar dressing
function $B$ and 
the transverse pion decay constant $f_s$.
In $\Sigma^B_{\text{YM}}$ and $\Sigma^B_{\pi,\sigma}$ the
scalar dressing $B$ occurs in the non-perturbative
quark propagator and in
the Bethe-Salpeter amplitude. The pion decay constant $f_s$
only occurs in $\Sigma^B_{\pi,\sigma}$ in the pion 
velocity $u=f_s/f_t $.
We stress that the scalar dressing function in the denominator
of the quark propagator can safely be ignored for our analysis,
since $B$ is very small close to $T_c$ and therefore
dominated by the large, non-vanishing fermionic Matsubara 
frequencies in the denominator. 

For the first contribution we find $\Sigma^B_{\text{YM}}(t) \sim B(t)$,
since only $B(\vq,\oq)$ in the integral kernel of
$\Sigma^B_{\text{YM}}$ scales with reduced 
temperature\footnote{Here we already use that the Yang-Mills vertex
dressing and the gluon dressing function are independent of $t$.
The latter assumption can be justified by a scaling analysis
of the quark-loop contribution to the gluon propagator along the lines
followed here. One consistently obtains scaling with $t$ proportional 
to $B^2/u^2 \sim const$, cf. Eq.~(\ref{scal_ana}). For the Yang-Mills
part of the vertex one can show that close to $T_c$ diagrammatic 
contributions containing only dressing functions $A,C$ dominate and
consequently this part of the vertex also does not scale.}.
Therefore self-consistency is trivially fulfilled and no constraint 
for $x$ can be derived from this. On the other hand,
the projection of the self-energies
$\Sigma^B_{\pi,\sigma}$ yields
\be
B(t)\sim\Sigma^B_{\pi,\sigma}(t)\sim \frac{B^3(t)}{{f^2_s(t)}}
+c\,\frac{B^5(t)}{{f^2_s(t)}}\,, \label{eq:Bsim}
\ee
where $c$ is a dimensionful constant with respect to $t$.
The numerator is obtained from
the fact that the trace over Dirac matrices is
only non-vanishing for an even number of matrices.
The denominator stems from the
zero frequency meson propagator with $\op^2=0$. 
For small $B$, i.e. close to $T_c$, the cubic term
is leading. With the scaling of the transverse
pion decay constant given in Eq.~(\ref{scal1})
we find
\be
t^x \sim \frac{t^{3x}}{t^\nu}\Rightarrow x=\nu/2\,. \label{scal_ana}
\ee
Thus our analysis suggests that in case of a 
second order phase transition of $N_f=2$ QCD
the quark scalar dressing function should be found
to behave like $t^{\nu/2}$ with $\nu\approx 0.73$.
Our result is supported by the fact that this solution is not only
consistent with the quark propagator Dyson-Schwinger equation but also
with the known relations for the quark condensate (\ref{eq:mscaling})
and for the transverse pion decay constant (\ref{scal1}).
To see this we recall that the quark condensate can be
obtained from the full quark propagator by
\begin{equation}
  \langle\bar{\psi}\psi \rangle = 
 Z_2\, N_c\, T\sum_{n_p}\int^{\Lambda}\frac{d^3p}{(2\pi)^3}\,
 \sigma_B(\vp,\op)\,\label{eq:QK}
\end{equation}
with $\sigma_B(\vp,\op)=B/(\op^2\,C^2(\vp,\op)+\vp^2 A^2(\vp,\op)+B^2(\vp,\op))$.
Therefore 
\beq
\langle\bar{\psi}\psi \rangle(t) \sim B(t) \sim t^{\nu/2}
\eeq 
consistent with (\ref{eq:mscaling})
in the approximation of vanishing anomalous dimension.
For the pion decay constant we conclude by inserting $B\sim t^{\nu/2}$
in Eq.~(\ref{pi_decay}) and evaluation for $P=(\vp,0)\rightarrow 0$:
\begin{equation}
B\sim t^{\nu/2}\Rightarrow {f_s}^2\sim t^{\nu}\,.
\end{equation}
Hence, we obtain a closed system in terms of scaling with 
reduced temperature $t$.

The above analysis relies on self-consistency of the quark-DSE.
Given a non-trivial scaling behavior of $f_s$, we find scaling
of the scalar part of the quark propagator, which is induced
via the pion dispersion relation. We therefore identified the 
mechanism for non-trivial scaling of the quark. However,
the emergence of non-mean-field critical exponents in $f_s$ 
is not explained. This issue needs to be further clarified in 
future studies. In the following we verify the analytical findings
of this section also in a numerical treatment of the quark-DSE.

\section{Numerical solutions of the DSE}

In the previous section we presented a self-consistent
approximation for the quark Dyson-Schwinger equation including
meson back-coupling effects.
This constitutes a two loop expression whose
numerical computation is still non-standard
and beyond the scope of this work.
Therefore and in order to avoid technical problems arising
in such a computation we simplify the expression further.
This is done similar to Refs.~\cite{Fischer:2008sp}.
We first give the details of this approximation and then
check the analysis of the previous section by setting a power law for
the decay constant $f_s$. 

\subsection{Truncation scheme \label{sec:trunc}}

\begin{figure}[t]
\centerline{\epsfig{file=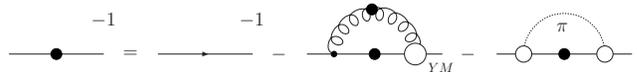,width=\columnwidth}}
\caption{The approximated Schwinger-Dyson equation
for the quark propagator.
\label{fig:quarkdse-vert2}}
\end{figure}

In order to further simplify the diagram in the lower line of
Fig.~\ref{fig:quarkdse} note that one of the loops involves 
two bare quark-gluon vertices, a dressed gluon propagator and 
a full Bethe-Salpeter vertex. If the gluon propagator and the 
two quark-gluon vertices would be replaced by a full Bethe-Salpeter 
kernel this loop would exactly match a Bethe-Salpeter equation. 
From this we deduce that the leading part of this loop is given 
by a term proportional to $Z_2 \gamma_5 \tau^i$ in the pseudoscalar 
meson sector and $Z_2 \tau^i$ for a scalar meson. In the vacuum
it has been shown in Ref.~\cite{Fischer:2008wy} that good 
results for meson phenomenology can already be obtained if the
proportionality factor is set to one. Here such a simple truncation
is not possible, since it would destroy the scaling properties 
close to $T_c$ analyzed in the previous section. Instead, we need 
to replace the loops by $Z_2 \gamma_5 \tau^i$ and $Z_2 \tau^i$, 
each modified with their respective quark-meson couplings 
$\Gamma_{\pi,\sigma}$. The resulting approximated quark Dyson-Schwinger 
equation, displayed in Fig.~\ref{fig:quarkdse-vert2}, has the same scaling
properties as its two-loop counterpart in Fig.~\ref{fig:quarkdse}.
Furthermore it is sufficient for the scaling analysis to consider only
the zeroth Matsubara frequency of the mesons and choose $m_\pi = m_\sigma=0$.
The corresponding equation is then given by
\begin{widetext}
\begin{eqnarray}
  S^{-1}(\vp,\op) &=& Z_{2} S^{-1}_{0}(\vp,\op)
  + g^{2}C_{F}Z_{1F} T \intq \,  \gamma_{\mu} S(\vq,\oq) 
  \Gamma_{\nu}^{\text{YM}}(\vq,\oq,\vp,\op)
  D_{\mu\nu}(\vk,\ok)   \nonumber\\
  &&\hspace*{0cm} -T \int \frac{d^3q}{(2\pi)^3} \, 
  \bigg(2\,\Gamma_{\pi}^T(\vp,\op) S(\vq,\op) 
  \Gamma_{\pi}(\vq,\op) D_\pi(\vk,0)+\Gamma_{\sigma}^T(\vp,\op) S(\vq,\op) 
  \Gamma_{\sigma}(\vq,\op) D_\sigma(\vk,0)\bigg)\;.
  \label{eq:DSE_detail}
\end{eqnarray}
\end{widetext}

As an aside, note that this truncation also fulfills the 
Mermin-Wagner theorem: Using Eq.~(\ref{eq:BSA1}) for the 
Bethe-Salpeter amplitude we find that in case of spontaneous 
symmetry breaking the Goldstone boson propagator $D_\pi(\vk,0)$ 
would lead to an infrared singular integral in 
Eq.~(\ref{eq:DSE_detail}) for $d\leq 2$ space dimensions. 
The singularity would shift $B(\vp,\omega_p)\rightarrow -\infty$.
Therefore in $d\leq 2$ space dimensions only the symmetric phase 
with $B(\vp,\omega_p)=0$ can be realized in agreement with 
Mermin-Wagner.

To implement Eq.~(\ref{eq:DSE_detail}) it remains to 
specify the input for the Yang-Mills part of the quark-gluon 
interaction and the gluon propagator. Since we are looking at 
critical scaling related to universality we expect that the 
details of both are not crucial for our studies. We have checked 
this by using two different approaches: On the one hand we used 
a simple rainbow-ladder like phenomenological model, which is 
detailed in appendix \ref{app:model}. On the other hand we 
followed Ref.\cite{Fischer:2011mz} and used quenched lattice 
data for the temperature dependent gluon propagator together 
with an explicit back-coupling of the quarks onto the gluon
propagator as input, see appendix \ref{app:model} for more 
details. Indeed we have found that the results for the critical
scaling behavior close to $T_c$ are similar in both approaches, 
as expected. In the next section we therefore concentrate 
on one of the two approximation schemes and present results 
for the more advanced truncation with temperature dependent 
gluon only.

\subsection{Results: scaling of quark dressing and condensate 
\label{sec:o4numerics}}

\begin{figure}[t]
\centerline{\epsfig{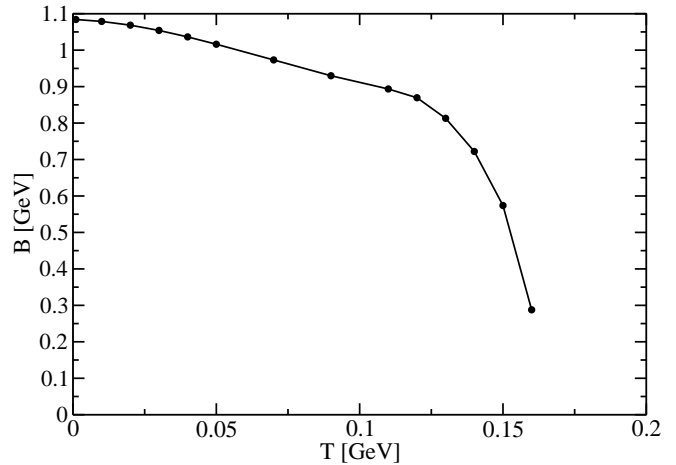}}
\caption{The variation of the scalar quark dressing function $B(\vp=0,\op=\pi T)$
with temperature not too close to the chiral critical temperature $T_c$.
\label{fig:scaling0}}
\end{figure}

In order to study scaling close to the critical temperature we
first have to determine $T_c$. This can be done either from the
behavior of the scalar quark dressing function $B$ or, equivalently,
from the chiral condensate. Our results for the former are shown
in Fig.~\ref{fig:scaling0}. We observe a reduction for
$B(\vp=0,\op=\pi T,T)$ with temperature which results in a phase 
transition at $T_c = 162.5 \pm 2.5$ MeV, where the error reflect the
coarseness of our temperature grid. For small temperatures the 
details of this reduction do depend on the details of the interaction,
since the Yang-Mills part of the quark-gluon interaction is leading there.
For temperatures close to the critical one, effects from meson
back-coupling become more and more important due to the long range nature
of the meson dispersion relations. In Fig.~\ref{fig:scaling0} we
have been using all Matsubara frequencies in the pion propagators.
Neglecting all but the zeroth Matsubara frequency
(which is sufficient for scaling around $T_c$) and considering
also the sigma propagator which becomes important close to $T_c$ 
we obtain a similar 
picture with only slightly changed temperature $T_c = 157 \pm 2.5$ MeV. 
We use this value for the following scaling analysis.

As shown analytically in section \ref{sec:scaling} we obtain the correct 
$O(4)$-scaling in the quark-DSE provided we feed in the correct 
scaling properties of the spatial part of the pion decay constant 
$f_s$. In order to numerically test this scenario we choose the 
following scaling ansatz for $f_s$
\begin{equation}
 \tilde{f}_s(T) = \frac{f_s(T_0)}{(T_c-T_0)^{\nu/2}}(T_c-T)^{\nu/2}\,\label{eq:ansatz_decay}
\end{equation}
where $T_0=150$ MeV and $f_s(T_0)$ is the result for the decay 
constant obtained from Eq.~(\ref{pi_decay}). The critical temperature 
is chosen to be $T_c=162$ MeV. With this construction our numerical results 
smoothly connect with the ones for $f_s(T)$ at $T\leq150$ MeV. The exponent is
set to $\nu=0.73$ corresponding to the value of the $O(4)$ universality class.

\begin{figure}[t]
\centerline{\epsfig{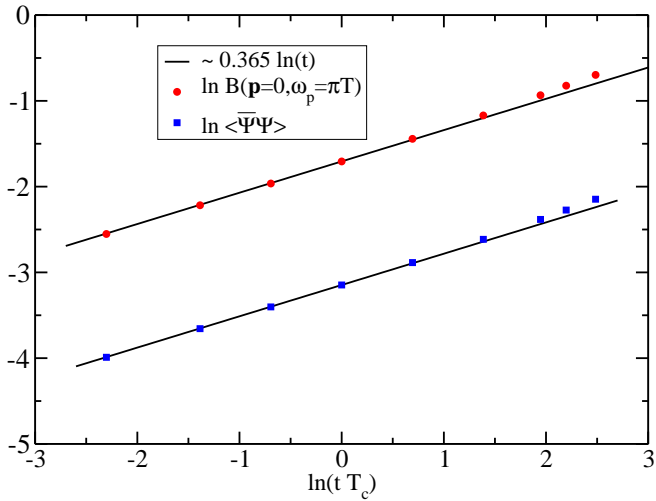}}
\caption{Scaling of the scalar quark dressing function $B(\vp=0,\op=\pi T)$
as well as the condensate with reduced temperature $t=(T_c-T)/T_c$. \label{fig:scaling1}}
\end{figure}

In figure \ref{fig:scaling1} the logarithm of the zeroth mode of the scalar 
dressing function $B$ (red circles) at zero momentum is shown as a function 
of $\ln(t\, T_c)$. In addition, the proportionality to $t^{\nu/2}$ is shown 
by the solid line.
For temperatures close enough to the critical one, $B(\vp=0,\op=\pi T)$ behaves 
exactly as it should according to our analytical scaling analysis
in section \ref{sec:scaling}. We also show results for the chiral condensate 
(blue squares) calculated from Eq.~(\ref{eq:QK}). They likewise follow such 
a scaling behavior as seen by the straight line $\sim t^{\nu/2}$. The
size of the scaling region can be inferred from the deviations from
scaling at large $t$ which set in at roughly $T_c-T \approx 10$MeV. 

Finally we investigated, whether the scaling of $f_s$ can be obtained
from its defining equation (\ref{pi_decay}). To this end we abandon the 
scaling ansatz (\ref{eq:ansatz_decay}) but calculate $f_s$ self-consistently
from Eq.~(\ref{pi_decay}). It turns out, that in this case we do
observe scaling around $T_c$, but the corresponding anomalous dimensions 
are the mean field ones. We attribute this result to the approximations
done in the pion sector, which reduce Eq.~(\ref{pi_decay}) to the
Pagels-Stokar form which may not be sufficiently accurate close to $T_c$.

\section{Conclusions and outlook \label{sec:con}}
In this paper we investigated the quark propagator close to
the second order chiral phase transition for $N_f=2$ massless
quarks in the symmetry broken phase.
We used the continuum formulation of the theory utilizing
the Dyson-Schwinger functional framework. In contrast to other methods
we have the ability to work in the chiral and thermodynamic limit
within this approach.
In order to take into account the relevant degrees of freedom 
we outlined a truncation scheme for the quark propagator
including meson backcoupling effects. In particular we studied
the effect of the Goldstone bosons, the pions, and the radial excitation,
the sigma meson.

Identifying the quantities of our truncation obeying known scaling
laws we analytically derived the scaling law of the scalar
dressing function $B$ with reduced temperature.
Our analysis reveals that the behavior of the pion velocity
$u$ respectively the pion decay constant $f_s$
is crucial for the scaling of the scalar dressing function
close to $T_c$. Furthermore within the employed approximation
(neglecting $\eta$) the result is consistent with the known
scaling relations for the chiral condensate
and the decay constant. 

In addition we also considered the quark Dyson-Schwinger equation
numerically. Here we could verify our analytical findings also 
numerically: using an appropriate scaling ansatz for the pion
decay constant we showed that the scalar quark dressing function
and the chiral condensate scale with $O(4)$-critical exponents
in a temperature region of about $10$ MeV below the chiral 
critical temperature. Using two different truncation schemes for
the Yang-Mills part of the interaction we explicitly verified, that 
chiral critical scaling is independent of the details of the
gluon propagator and the Yang-Mills part of the quark-gluon vertex,
as expected from universality. 

Finally we performed a first attempt to self-consistently close 
the system by determining the pion decay constant from the Pagel-Stokar 
formula. This, however, turned out not to be sufficient, since the
resulting critical exponents returned to their mean field values.
We interpret this a a failure of the Pagel-Stokar approximation 
close to the chiral phase transition. Going beyond this approximation
will be a task for future work.

\vspace*{3mm}
{\bf Acknowledgments}\\
We thank J\"urgen Berges, Jan Luecker, Bernd-Jochen Schaefer, 
and Lorenz von Smekal for useful discussions and Bernd-Jochen Schaefer
for a critical reading of the manuscript. This work has been 
supported by the Helmholtz-University Young Investigator Grant 
No.~VH-NG-332 and by the Helmholtz International Center for FAIR 
within the LOEWE program of the State of Hesse. \\[1mm]

\begin{appendix}

\section{Two truncations of the Yang-Mills part of the quark-DSE\label{app:model}}

In the main part of this work we numerically investigate a combination of 
purely gluonic effects and meson back-coupling effects onto the quark propagator.
Since the Yang-Mills part of the quark-gluon interaction does not take part
in the critical $O(4)$-scaling close to $T_c$, it is sufficient for
our purpose to approximate it in the simple fashion of a rainbow-ladder like
model \cite{Alkofer:2008et}. For the Yang-Mills part of the quark-gluon vertex 
$\Gamma^{\text{YM}}_\mu = \gamma_\mu \Gamma^{\text{YM}}(z)$ with the squared gluon
momentum $z=k^2=\omega_k^2 + \vec{k}^2=(\oq-\op)^2+(\vq-\vp)^2$ we use the following model ansatz 
\begin{align}
\Gamma^{\text{YM}}(z)=& \left(\frac{z}{z+
d_2}\right)^{-1/2-\kappa}\Bigg( \frac{ d_1}{d_2+z}+
\frac{ z d_3}{d_2^2+\left(z- d_2\right)^2}
\nonumber\\[-1.2mm]
+&\frac{z}{ \Lambda^2_{\text{QCD}}+z}\times\nonumber\\[-1.2mm]
&\left[\frac{4\pi}{\beta_0\alpha_\mu}
\left(\frac{1}{\log\left(z/{\Lambda^2_{\text{QCD}}}\right)}-\frac{\Lambda^2_{\text{QCD}}}{z-\Lambda^2_{\text{QCD}}}\right)\right]^{-2\delta}
	    \Bigg)\nonumber \\
\label{eqn:l1param}
\end{align}
with parameters $d_1=1.45 \mbox{GeV}^2$, $d_2=0.1 \mbox{GeV}^2$ and 
$d_3=3.95 \mbox{GeV}^2$.
Furthermore the QCD scale is $\Lambda_{\text{QCD}}=0.520$ GeV,
$\alpha_\mu=0.2$ and $\kappa \approx 0.595 $. Determined from
perturbation theory we find $\delta = -9 N_c/(44 N_c-8 N_f)$
 and $\beta_0=(11 N_c-2 N_f)/3$.
The gluon propagator dressing functions are approximated 
by a fit function to the zero temperature results
\begin{equation}
 Z(z)=Z_T(z)=Z_L(z)\nonumber\\
\end{equation}
\begin{eqnarray}
Z(z)&=&\left(\frac{z}{z+\Lambda^2_{QCD}}\right)^{2\kappa}\;\left(\frac{1}{\alpha_\mu}\,
\left[\frac{\alpha_0}{1+z/\Lambda^2_{QCD}} \right.\right.\nonumber\\
&&\hspace{-1.2cm}\left.\left.+\frac{4\pi}{\beta_0}\frac{z}{z+\Lambda^2_{QCD}}\left(\frac{1}{\ln(z/\Lambda^2_{QCD})}
-\frac{1}{z/\Lambda^2_{QCD}-1}\right)\right]\right)^{-\gamma}\nonumber\\\,
\;\label{eq:glueRL}
\end{eqnarray}
where $\beta_0=(11 N_c-2 N_f)/3$.
The different temperature dependence of the longitudinal
and transverse dressing functions is neglected. 
\begin{figure}[b]
\centerline{\includegraphics[width=\columnwidth]{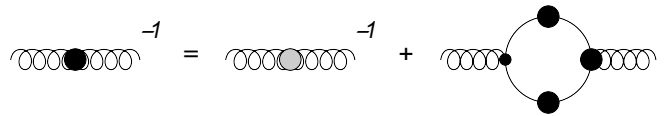}}
\caption{The DSE for the gluon propagator. Filled circles denote 
dressed propagators and vertices. The shaded circle denotes the 
quenched gluon.}
\label{fig:DSE}
\end{figure}

As an alternative we also employ the much more complex interaction
developed in \cite{Fischer:2011mz}. There, quenched lattice data for the 
temperature dependent gluon propagator have been combined with the 
quark loop polarization diagram to obtain a numerical expression for 
the unquenched gluon propagator with $N_f=2$ quarks back-coupled. 
Unfortunately, there is no reliable information on the details of the
quark-gluon vertex at finite temperature, so also in this approach one 
has to resort to a model for the vertex. Here we choose
\begin{widetext}
\beqa \label{vertexfit}
\Gamma_\nu(q,k,p) = \widetilde{Z}_{3}\gamma_\nu\left( 							
\frac{d_1}{d_2+q^2} 			
 + \frac{q^2}{\Lambda^2+q^2}
\left(\frac{\beta_0 \alpha(\mu)\ln[q^2/\Lambda^2+1]}{4\pi}\right)^{2\delta}\right) \,,
\eeqa 
\end{widetext}
where $q=(\vq,\oq)$ denotes the gluon momentum and $p=(\vp,\op)$, $k=(\vk,\ok)$ the
quark and antiquark momenta, respectively. Furthermore 
$2\delta = -18 N_c/(44 N_c-8 N_f)$ is the anomalous dimension 
of the vertex and the parameters are $d_1=15 \,\mbox{GeV}^2$,$d_2=0.5 \,\mbox{GeV}^2$,
$\Lambda = 1.4$ GeV and we renormalize at $\alpha(\mu)=0.3$.

The numerical results in section \ref{sec:o4numerics} of this work have been obtained
using the second, more complicated truncation of the Yang-Mills part of the
quark-gluon interaction. However, we wish to emphasize once more that the
results using the simpler rainbow-ladder form, (\ref{eqn:l1param}) and 
(\ref{eq:glueRL}), are similar. Critical scaling of the chiral order parameters 
is universal and thus independent of the details of the Yang-Mills part of the 
quark-gluon interaction. In particular this is true for the deep infrared behavior:
whereas the rainbow ladder approximation (\ref{eqn:l1param}) and (\ref{eq:glueRL}) 
is constructed to adhere to the Yang-Mills 'scaling' scenario (in the terms of
Ref.~\cite{Fischer:2008uz}), the more complicated construction (\ref{vertexfit}) 
together with the temperature dependent lattice gluon propagator of 
Ref.~\cite{Fischer:2011mz} behave according to the 'decoupling' scenario 
(again in terms of Ref.~\cite{Fischer:2008uz}). Although the deep infrared 
behavior of these two scenarios is very different, it does not affect the 
chiral critical scaling properties of the systems at $T_c$, in agreement 
with previous findings \cite{Fischer:2009wc}.

\end{appendix}

\end{document}